\documentclass[aps,prd,preprintnumbers,superscriptaddress,nofootinbib,showpacs,twocolumn]{revtex4}%
\usepackage[dvipdfmx]{graphicx}
\usepackage{bm,latexsym,amsmath,amssymb,amsfonts,mathrsfs}
\usepackage{color}
\input{colordvi.tex}
\allowdisplaybreaks[1]
\usepackage[dvipdfmx]{hyperref}
\usepackage{url}
\hypersetup{
    colorlinks=true,
    citecolor=cyan,
}
\newcommand*{\D}{{\rm d}}
\newcommand*{\mpl}{M_{\rm Pl}}

\definecolor{pn}{rgb}{1.0,0.2,0.8}

\begin{document}

\title{Primordial non-Gaussianities of
scalar and tensor perturbations in general bounce cosmology:
Evading the no-go theorem}

\author{Shingo~Akama}
\email[Email: ]{s.akama"at"rikkyo.ac.jp}
\affiliation{Department of Physics, Rikkyo University, Toshima, Tokyo 171-8501, Japan
}
\author{Shin'ichi Hirano}
\email[Email: ]{s.hirano"at"rikkyo.ac.jp}
\affiliation{Department of Physics, Rikkyo University, Toshima, Tokyo 171-8501, Japan
}
\author{Tsutomu~Kobayashi}
\email[Email: ]{tsutomu"at"rikkyo.ac.jp}
\affiliation{Department of Physics, Rikkyo University, Toshima, Tokyo 171-8501, Japan
}

\begin{abstract}
It has been pointed out that matter bounce cosmology driven by a k-essence field cannot satisfy simultaneously the observational bounds on the tensor-to-scalar ratio and non-Gaussianity of the curvature perturbation. 
In this paper, we show that this is not the case in more general scalar-tensor theories.
To do so, we evaluate the power spectra and the bispectra of scalar and tensor perturbations on a general contracting background in the Horndeski theory.
We then discuss how one can discriminate contracting models from inflation based on non-Gaussian signatures of tensor perturbations.
\end{abstract}

\preprint{RUP-19-23}
\maketitle

\section{Introduction}
Although it is definite that
inflation~\cite{Guth:1980zm,Starobinsky:1980te,Sato:1980yn} is the most
 successful early universe model,
it is inevitably plagued by the initial singularity problem~\cite{Borde:1996pt}.
Motivated by this,
alternative scenarios which do not suffer from this problem
have also been explored
(see, e.g.,~\cite{Battefeld:2014uga} for a review).
Non-singular cosmology has its own difficulty
regarding gradient instabilities when constructed within
second-order scalar-tensor theories~\cite{Libanov:2016kfc,Kobayashi:2016xpl,Cai:2016thi,Creminelli:2016zwa,Akama:2017jsa},
but its resolution has been proposed
in the context of higher-order scalar-tensor theories~\cite{Cai:2016thi,Creminelli:2016zwa,Cai:2017tku,Cai:2017dyi,%
Kolevatov:2017voe,Ye:2019frg,Ye:2019sth}.
It is also important to discuss the validity of
non-singular alternatives from the viewpoint of cosmological observations.

For example, a matter-dominated contracting (or bounce) universe
can be mimicked by a canonical scalar field
and this model can generate
a scale-invariant curvature
perturbations~\cite{Wands:1998yp,Finelli:2001sr,Quintin:2015rta}.
However,
this model yields a too large tensor-to-scalar ratio and
thus is excluded~\cite{Quintin:2015rta} (see, however, Refs.~\cite{Raveendran:2017vfx,Raveendran:2018why}).
One may use a k-essence field
to reduce the tensor-to-scalar ratio
by taking a small sound speed,
but then this in turn enhances the production of non-Gaussianity,
making the model inconsistent with observations~\cite{Li:2016xjb}.
At this stage, it is not evident whether or not this
``no-go theorem''
holds in more general scalar-tensor theories.

The purpose of the present paper is
clarifying to what extent the previous no-go theorem
(which was formulated in the context of a k-essence field
minimally coupled to gravity as
an extension of Ref.~\cite{Quintin:2015rta})
holds in more general setups. To do so, we consider a
general power-law contracting universe in the Horndeski theory~\cite{Horndeski:1974wa},
the most general second-order scalar-tensor theory,
and evaluate the power spectra and the bispectra
of scalar and tensor perturbations generated
during the contracting phase.
Throughout the paper we assume that the statistical nature
of these primordial perturbations does not change
during the subsequent bouncing and expanding phases.
(In some cases in matter bounce cosmology, this has been justified.
See, e.g., Ref.~\cite{Gao:2009wn}.)
In calculating tensor non-Gaussianity
we explore peculiar signatures of a contracting phase
as compared to inflation, and show that
the two scenarios can potentially be distinguishable
due to the non-Gaussian amplitudes and shapes.

This paper is organized as follows.
In the next section, we introduce our setup of the
general contracting cosmological background.
In Sec.~III, we evaluate the power spectra for
curvature and tensor perturbations, and derive the conditions
under which they are scale-invariant.
In Sec.~IV, we calculate primordial non-Gaussianities of
curvature and tensor perturbations,
and investigate whether
a small tensor-to scalar ratio and small scalar non-Gaussianity are compatible or not
in the Horndeski theory.
We also discuss how one can distinguish bounce cosmology with
inflation based on tensor non-Gaussianity.
The conclusion of this paper is drawn in Sec. V.

\section{Setup}

We begin with a spatially flat
Friedmann-Lema\^{i}tre-Robertson-Walker
(FLRW) metric
\begin{align}
{\D}s^2=-{\D}t^2+a^2(t)\delta_{ij}{\D}x^i{\D}x^j,
\end{align}
where the scale factor describes a contracting phase,
\begin{align}
a=\left(\frac{-t}{-t_b}\right)^{n}=\left(\frac{-\eta}{-\eta_b}\right)^{n/(1-n)}
\quad  \left(0<n<1\right), \label{scaf}
\end{align}
with $\D\eta=\D{t}/a$. Here, we denoted
the time at the end of the contracting phase
as $t_b(<0)$ and $\eta_b(<0)$, and we normalized the scale factor so that $a(t_b)=1=a(\eta_b)$.
The two time coordinates are related with
\begin{align}
-\eta=\frac{(-t_b)^n}{1-n}(-t)^{1-n},
\end{align}
where $t$ and $\eta$ coordinates run from $-\infty$ to $t_b$ and $\eta_b$, respectively. 
In this paper, we do not assume $n$ to take any particular value,
so that our setup includes models other than the familiar
matter bounce scenario~\cite{Brandenberger:2012zb}. Note, however, that it will
turn out that models with different $n$ are related to each other
via conformal transformation (see Sec.~\ref{subsec:ct}).

We work with the Horndeski action which is given by
\begin{align}
S=\int{\D^4x}\sqrt{-g}\mathcal{L},
\end{align}
with
\begin{align}
\mathcal{L}&=G_2(\phi,X)-G_3(\phi,X)\Box\phi
+G_4(\phi,X)R
\notag \\ &\quad
+G_{4X}\left[(\Box\phi)^2-(\nabla_{\mu}\phi\nabla_{\nu}\phi)^2\right]
\notag \\ &\quad
+G_5(\phi,X)G_{\mu\nu}\nabla^{\mu}\nabla^{\nu}\phi
-\frac{G_{5X}}{6}\bigl[(\Box\phi)^3
\notag \\ &\quad
-3\Box\phi(\nabla_{\mu}\nabla_{\nu}\phi)^2+2(\nabla_{\mu}\nabla_{\nu}\phi)^3\bigr],
\end{align}
where $X:=-g^{\mu\nu}\nabla_\mu\phi\nabla_\nu\phi/2$ and
 $\partial G/\partial X$ is denoted by $G_{X}$.
This action gives the most general second-order scalar-tensor theory,
and hence a vast class of contracting scenarios
reside within this theory.
Therefore, the Horndeski theory is
adequate for studying generic properties of
cosmological perturbations from contracting models.
Note, however, that nonsingular cosmological solutions
suffer from gradient instabilities if the entire history of the
universe were described by the Horndeski theory~\cite{Libanov:2016kfc,Kobayashi:2016xpl,Cai:2016thi,Creminelli:2016zwa,Akama:2017jsa}.
We circumvent this issue by assuming that beyond-Horndeski operators
come into play at some moment, but at least the contracting phase
we are focusing on is assumed to be described by the Horndeski theory.

The Friedmann and evolution equations are written,
respectively,
in the form
\begin{align}
\mathcal{E}:=\sum_{i=2}^5\mathcal{E}_i=0,
\quad
\mathcal{P}:=\sum_{i=2}^5\mathcal{P}_i=0,
\end{align}
where
$\mathcal{E}_i=\mathcal{E}_i(H,\phi,\dot \phi)$ and
$\mathcal{P}_i=\mathcal{P}_i(H,\dot H,\phi,\dot \phi,\ddot \phi)$
come from the variation of the action involving $G_i$,
whose explicit expressions are given in Appendix~\ref{App.A}.
Here a dot stands for differentiation with respect to $t$
and $H:=\dot a/a$. In this paper, we do not
consider any concrete background models, but just assume that
each term in the background equations scales as
\begin{align}
\mathcal{E}_i,\ \mathcal{P}_i\sim(-t)^{2\alpha}, \label{BG}
\end{align}
where $\alpha$ is a constant to be specified below.
The impact of spatial curvature and anisotropies is discussed
in Appendix~\ref{App.B}.

\section{Scale-invariant power spectra}

The perturbed metric in the unitary gauge, $\delta\phi(t,{\bf x})=0$, is written as
\begin{align}
{\D}s^2=-N^2{\D}t^2+g_{ij}\left({\D}x^i+N^i{\D}t\right)\left({\D}x^j+N^j{\D}t\right), \label{perturb-met}
\end{align}
where
\begin{align}
& N=1+\delta{n},\quad N_i=\partial_i\chi,\quad
g_{ij}=a^2e^{2\zeta}(e^h)_{ij},\label{pertmet}
\\
&(e^h)_{ij}:=\delta_{ij}+h_{ij}+\frac{1}{2}h_{ik}h^k_j+\frac{1}{6}h_{ik}h^k_lh^l_j
+\cdots.
\end{align}
As has been done in Ref.~\cite{Kobayashi:2011nu},
one expands the action to second order in perturbations
and removes the auxiliary variables $\delta n$ and $\chi$.
The resultant
quadratic actions for the curvature perturbation $\zeta$ and the
tensor perturbations $h_{ij}$
in the Horndeski theory are written, respectively, as
\begin{align}
S^{(2)}_{\zeta}&=\int{\D t}\D^3xa^3\left[\mathcal{G}_S\dot{\zeta}^2-\frac{\mathcal{F}_S}{a^2}(\partial_i\zeta)^2\right],\\
S^{(2)}_T&=\frac{1}{8}\int{\D t}\D^3xa^3\left[\mathcal{G}_T{\dot h_{ij}}^2-\frac{\mathcal{F}_T}{a^2}(\partial_k h_{ij})^2\right],
 \end{align}
where
 \begin{align}
\mathcal{G}_T&=2\left[G_4-2XG_{4X}-X\left(H\dot\phi{G_{5X}}-G_{5\phi}\right)\right], \label{GT}\\
\mathcal{F}_T&=2\left[G_4-X\left(\ddot{\phi}G_{5X}+G_{5\phi}\right)\right], \label{FT}\\
\mathcal{G}_S&=\mathcal{G}_T\left(\frac{\mathcal{G}_T\Sigma}{\Theta^2}+3\right), \label{GS}\\
\mathcal{F}_S&=\frac{1}{a}\frac{\rm d}{\rm dt}\left(\frac{a\mathcal{G}_T^2}{\Theta}\right)-\mathcal{F}_T, \label{FS}
\end{align}
with
\begin{align}
\Sigma&=X\frac{\partial\mathcal{E}}{\partial X}+\frac{H}{2}\frac{\partial\mathcal{E}}{\partial H}, \label{Sigma}\\
\Theta&=-\frac{1}{6}\frac{\partial\mathcal{E}}{\partial H}. \label{Theta}
\end{align}
(The explicit expressions for $\Theta$ and $\Sigma$ are given in Appendix~\ref{App.C}.)
As inferred from Eqs.~(\ref{BG}),~(\ref{Sigma}), and~(\ref{Theta}),
it is natural to assume that
$\Sigma\sim(-t)^{2\alpha}$ and $\Theta\sim(-t)^{2\alpha+1}$.
In addition,
it can be seen that
${\cal G}_T, {\cal F}_T\sim {\cal E}_4/H^2, {\cal E}_5/H^2,{\cal P}_4/H^2, {\cal P}_5/H^2$.
These imply
\begin{align}
&\mathcal{G}_T,\ \mathcal{F}_T,\ \mathcal{G}_S,\ \mathcal{F}_S\sim(-t)^{2(\alpha+1)}\propto(-\eta)^{2(\alpha+1)/(1-n)}. \label{GF}
\end{align}
Under these assumptions,
the propagation speed of the curvature perturbation, $c_s^2=\mathcal{F}_S/\mathcal{G}_S$,
and that of the tensor perturbations, $c_t^2=\mathcal{F}_T/\mathcal{G}_T$,
are constant. Note that only $\alpha=-1$ is possible
if $\phi$ is minimally coupled to gravity.

Let us move to derive a relation between $\alpha$ and $n$
by imposing that the primordial curvature and tensor perturbations
have scale-invariant power spectra.

\subsection{Curvature Perturbation}
We expand and quantize the curvature perturbation as
\begin{align}
\zeta(t,{\bf x})&=\int\frac{\D^3k}{(2\pi)^3}\hat\zeta(t,{\bf k})e^{i{\bf k}\cdot{\bf x}},\\
                              &=\int\frac{\D^3k}{(2\pi)^3}\left[\zeta_{\bf k}(t)\hat{a}_{\bf k}+{\zeta}^*_{-\bf k}(t)\hat{a}^{\dagger}_{-\bf k}\right]e^{i{\bf k}\cdot{\bf x}},
\end{align}
where the commutation relations between the creation and annihilation operators are standard ones,
\begin{align}
\left[\hat{a}_{\bf k},\hat{a}^{\dagger}_{{\bf k}'}\right]&=(2\pi)^3\delta\left({\bf k}+{\bf k}'\right),\\
{\rm{others}}&=0.
\end{align}
The mode function $u_{\bf k}(\eta)$ of the canonically normalized perturbation,
$u_{\bf k}=\sqrt{2}a(\mathcal{F}_S\mathcal{G}_S)^{1/4}\zeta_{\bf k}$, obeys
\begin{align}
u_{\bf k}''+\left[c_s^2k^2-\frac{1}{\eta^2}\left(\nu_s^2-\frac{1}{4}\right)\right]u_{\bf k}=0,
\end{align}
where a prime denotes differentiation with respect to $\eta$ and
\begin{align}
\nu_s:=\frac{-1-3n-2\alpha}{2(1-n)}.
\end{align}
The positive frequency solution is then given by
\begin{align}
\zeta_{\bf k}=\frac{1}{\sqrt{2}a({\cal F}_S{\cal G}_S)^{1/4}}
\cdot \frac{\sqrt{\pi}}{2}\sqrt{-c_s\eta}H_{\nu_s}^{(1)}(-c_sk\eta),
\end{align}
where $H_{\nu}^{(1)}$ is
the Hankel function of the first kind. Here we chose the
initial condition as
\begin{align}
\lim_{\eta\to-\infty}u_{\bf k}=\frac{1}{\sqrt{2k}}e^{-ic_sk\eta}.
\end{align}
The power spectrum of the curvature perturbation is defined by
\begin{align}
\langle{ \hat\zeta({\bf k})\hat\zeta({\bf k}')}\rangle=(2\pi)^3\delta\left({\bf k}+{\bf k}'\right)\frac{2\pi^2}{k^3}\mathcal{P}_\zeta(k),
\end{align}
and therefore
\begin{align}
{\cal P}_\zeta \propto k^{3-2|\nu_s|}.
\end{align}
The spectral index is thus given by
\begin{align}
n_s-1=3-2|\nu_s|.
\end{align}
Let us focus on the exactly scale-invariant spectrum,
which corresponds to
\begin{align}
\nu_s=\frac{3}{2}\quad&\Rightarrow\quad
\alpha=-2,\\
\nu_s=-\frac{3}{2}\quad&\Rightarrow\quad
\alpha =1-3n .\label{nus-32}
\end{align}
On superhorizon scales, $c_sk|\eta|\ll 1$, we have
$\zeta_\mathbf{k}\propto |\eta|^{\nu_s-|\nu_s|}$.
Therefore,
the perturbations freeze out on superhorizon scales
in the former case (as in the inflationary universe),
while they grow as $\zeta_\mathbf{k}\propto |\eta|^{-3}$
in the latter case (as in the contracting universe).
In this paper, we consider the growing superhorizon perturbations
having a scale-invariant spectrum, which is a
characteristic feature of contracting models.
Note that the Planck results~\cite{Akrami:2018odb} require
a slightly red tilted spectrum, $n_s\simeq0.96$.
This can be obtained by slightly detuning the relation~\eqref{nus-32}
between $n$ and $\alpha$, though for simplicity in this paper
we only consider
the exactly scale-invariant case.

Taking $\alpha=1-3n$,
the scale-invariant power spectrum can now be derived as
\begin{align}
\mathcal{P}_\zeta=
\left.\frac{1}{8\pi^2}\frac{1}{\mathcal{F}_{S}c_s}\frac{1}{\eta^2}\right|_{t=t_b}
=\left.\frac{1}{8\pi^2}\left(1-\frac{1}{n}\right)^2\frac{H^2}{\mathcal{F}_{S}c_s}
\right|_{t=t_b}, \label{PScur}
\end{align}
where the time-dependent quantities are
evaluated at the end of the contracting phase.

\subsection{Tensor Perturbations}

The tensor perturbations can be expanded and quantized as
\begin{align}
h_{ij}(t,{\bf x})&=
\int\frac{\D^3k}{(2\pi)^3}\hat{h}_{ij}(t,{\bf k})e^{i{\bf k}\cdot{\bf x}}\\
&=\sum_s\int\frac{\D^3k}{(2\pi)^3}\biggl[{h}^{(s)}_{\bf k}(t)\hat{a}^{(s)}_{\bf k}e^{i\bf{k}\cdot\bf{x}}e^{(s)}_{ij}(\bf k)\notag\\
&\quad\quad\quad\quad\quad\quad\quad+{h}^{(s)*}_{-{\bf k}}(t)
\hat{a}^{(s)\dagger}_{-{\bf k}}e^{(s)*}_{ij}(-{\bf k})\biggr]e^{i{\bf k}\cdot{\bf x}},
\end{align}
where the creation and annihilation operators satisfy the canonical commutation relations
\begin{align}
\left[\hat{a}^{(s)}_{\bf k},\hat{a}^{(s')\dagger}_{\bf k'}\right]&=(2\pi)^3\delta_{ss'}\delta\left({\bf k}+{\bf k'}\right),\\
{\rm{others}}&=0.
\end{align}
The two helicity modes are labeled
by $s=\pm$, 
and the basis $e^{(s)}_{ij}$ satisfies the transverse and traceless conditions, $\delta_{ij}e^{(s)}_{ij}({\bf k})=0=k^ie^{(s)}_{ij}(\bf k)$, and it is normalized as $e^{(s)}_{ij}({\bf k})e^{(s')*}_{ij}({\bf k})=\delta_{ss'}$.

The mode function $v^{(s)}_{\bf k}(\eta)$ of the canonically normalized
perturbations,
$v_{\bf k}^{(s)}=a(\mathcal{F}_T\mathcal{G}_T)^{1/4}h_{\bf k}^{(s)}/2$, obeys
\begin{align}
{v^{(s)}_{\bf k}}''+\left[c_t^2k^2-\frac{1}{\eta^2}\left(\nu_t^2-\frac{1}{4}\right)\right]v^{(s)}_{\bf k}=0,
\end{align}
where $\nu_t=\nu_s$.
The positive frequency solution is then given by
\begin{align}
h_{\bf k}^{(s)}=\frac{2}{a({\cal F}_T{\cal G}_T)^{1/4}}
\cdot \frac{\sqrt{\pi}}{2}\sqrt{-c_t\eta}H_{\nu_t}^{(1)}(-c_tk\eta),
\end{align}
where one can see that
\begin{align}
\lim_{\eta\to-\infty}v_{\bf k}^{(s)}=\frac{1}{\sqrt{2k}}e^{-ic_tk\eta}.
\end{align}

The behavior of the tensor perturbations is essentially the same as
that of $\zeta_\mathbf{k}$. For $\alpha=1-3n$ ($\nu_t=\nu_s=-3/2$),
$h_\mathbf{k}$ grows on superhorizon scales as $h_\mathbf{k}\propto |\eta|^{-3}$
and the tensor power spectrum is scale invariant.

Let us define ${\cal P}_{ij,kl}(\mathbf{k})$ by
\begin{align}
\langle{\hat h_{ij}({\bf k})\hat h_{kl}({\bf k}')}
\rangle=(2\pi)^3\delta\left({\bf k}+{\bf k}'\right)\mathcal{P}_{ij,kl}({\bf k}).
\end{align}
Then,
\begin{align}
\mathcal{P}_{ij,kl}({\bf k}):=\sum_s|h^{(s)}_{{\bf k}}(t)|^2\Pi_{ij,kl}({\bf k}),
\end{align}
with
\begin{align}
\Pi_{ij,kl}({\bf k}):=\sum_se^{(s)}_{ij}({\bf k})e^{(s)*}_{kl}({\bf k}),
\end{align}
and the tensor power spectrum is defined as
${\cal P}_h=(k^3/2\pi^2){\cal P}_{ij,ij}$.
For $\alpha=1-3n$,
we have the scale-invariant power spectrum
\begin{align}
 \mathcal{P}_h
 =\left.\frac{2}{\pi^2}\frac{1}{\mathcal{F}_{T}c_t}\frac{1}{\eta^2}\right|_{t=t_b}
 =\left.\frac{2}{\pi^2}\left(1-\frac{1}{n}\right)^2\frac{H^2}{\mathcal{F}_{T}c_t}
 \right|_{t=t_b},
 \end{align}
where time-dependent quantities are evaluated at $t=t_b$.

The tensor-to-scalar ratio is given by
\begin{align}
r=\frac{\mathcal{P}_h}{\mathcal{P}_\zeta}=
16\left.\frac{\mathcal{F}_{S}}{\mathcal{F}_{T}}\frac{c_s}{c_t}\right|_{t=t_b}, \label{ratio}
\end{align}
which is constrained as~\cite{Akrami:2018odb}
\begin{align}
r<0.064,\ &(95\%\ {\rm CL},\ Planck {\rm TT,TE,EE}\notag\\
                                                       &\ +{\rm lowE}+{\rm lensing}+{\rm BK14}). \label{const-r}
\end{align}

 For example,
 in the case of matter contracting models within the k-essence theory,
 we have $n=2/3$, $\alpha=-1$, $c_t=1$, and
 ${\cal F}_S=(3/2){\cal F}_T=\,$const.
 Therefore, the tensor-to-scalar ratio is
 \begin{align}
 r=24c_s, \label{r-kessence}
 \end{align}
 which can satisfy the upper bound on $r$ only
 for $c_s\ll 1$.
However, as argued in Ref.~\cite{Li:2016xjb}, small $c_s$
implies large scalar non-Gaussianity, and hence bounce models within
the k-essence theory are ruled out.
In the next section, we revisit this issue and study
whether or not upper bounds on the tensor-to-scalar ratio and
non-Gaussianity can be satisfied at the same time
in a wider class of theories.

\subsection{Conformal Frames}\label{subsec:ct}

At this stage it is instructive to perform
a conformal transformation and clarify the relation among
models with different $n$.

Let us consider a conformally related metric
\begin{align}
\widetilde{\D s}^2=\Omega^2(t)\left(-\D t^2+a^2\delta_{ij}\D x^i\D x^j\right),
\quad \Omega\propto (-t)^{\alpha+1}.
\end{align}
In this tilde frame, the time coordinate and the scale factor
are given respectively by
\begin{align}
\alpha=-2\quad&\Rightarrow\quad
-\tilde t \propto \ln(-t),
\quad
\tilde a \propto e^{\tilde{H}\tilde{t}},\\
\alpha\neq-2\quad&\Rightarrow\quad
-\tilde t \propto (-t)^{\alpha+2},
\quad
\tilde a \propto (-\tilde t)^{(n+\alpha+1)/(\alpha+2)}.
\end{align}
By inspecting the quadratic action for scalar and tensor perturbations
we see that in the tilde frame all the four coefficients
reduce to constants.

We find that the case of $\nu_s=\nu_t=3/2$ ($\alpha=-2$) can be regarded as de Sitter inflation (see, e.g., Ref.~\cite{Nandi:2019xag}).

In the case of $\nu_s=\nu_t=-3/2$ ($\alpha=1-3n$), we have
\begin{align}
\tilde a \propto (-\tilde t)^{2/3},
\end{align}
which describes a matter-dominated contracting universe.
Therefore, the dynamics of cosmological perturbations
in our contracting models (with general $n$) is equivalent to that
in the more familiar matter-dominated contracting model.
However, it should be emphasized that the magnitudes of the coefficients
in the perturbation action are still arbitrary even in the tilde frame.

\section{Primordial non-Gaussianities}

\subsection{Scalar Perturbations}
The three-point correlation function can be computed
by using the in-in formalism as
\begin{align}
&\langle\hat\zeta({\bf k}_1)
\hat \zeta({\bf k}_2)\hat \zeta({\bf k}_3)\rangle\notag\\
&=-i\int^{t_b}_{-\infty}\D t'\langle[\hat\zeta(t_b,{\bf k}_1)
\hat\zeta(t_b,{\bf k}_2)\hat\zeta(t_b,{\bf k}_3),H_{{\rm int}}(t')]\rangle, \label{InIn}
\end{align}
where
\begin{align}
H_{\rm int}=-\int \D^3x \mathcal{L}^{(3)}_{\zeta},
\end{align}
with ${\cal L}_\zeta^{(3)}$ being
the cubic Lagrangian of the curvature perturbation.
It can be written in the form~\cite{Gao:2011qe,DeFelice:2011uc,Gao:2012ib}
\begin{align}
{\cal L}_\zeta^{(3)}&=
a^3\mathcal{G}_S\biggl[\frac{\Lambda_1}{H}\dot\zeta^3+\Lambda_2\zeta\dot\zeta^2+\Lambda_3\zeta\frac{\left(\partial_i\zeta\right)^2}{a^2}\notag\\
&\quad  +\frac{\Lambda_4}{H^2}\dot\zeta^2\frac{\partial^2\zeta}{a^2}+\Lambda_5\dot\zeta\partial_i\zeta\partial_i\psi+\Lambda_6\partial^2\zeta\left(\partial_i\psi\right)^2\notag\\
& \quad +\frac{\Lambda_7}{H^2}\frac{1}{a^4}\left[\partial^2\zeta\left(\partial_i\zeta\right)^2-\zeta\partial_i\partial_j\left(\partial_i\zeta\partial_j\zeta\right)\right]\notag\\
& \quad +\frac{\Lambda_8}{H}\frac{1}{a^2}\left[\partial^2\zeta\partial_i\zeta\partial_i\psi-\zeta\partial_i\partial_j\left(\partial_i\zeta\partial_j\psi\right)\right]\biggr]\notag\\
&\quad +F(\zeta)E_S, \label{cubic-s}
\end{align}
where
$\psi:=\partial^{-2}\dot\zeta$ and
$\Lambda_i$ are dimensionless coefficients.
The complete form of the cubic Lagrangian is summarized in Appendix~\ref{App.C}.
Based on the scaling argument similar to that in the previous section,
it can be seen that the coefficients $\Lambda_i$ are constant.

The last term in Eq.~(\ref{cubic-s}) can be eliminated by means of a field redefinition
\begin{align}
\zeta\to\zeta-F(\zeta).
\end{align}
In Fourier space, this redefinition is equivalent to
\begin{widetext}
\begin{align}
\zeta({\bf k})&\to\zeta({\bf k})-\frac{3(1-n)}{n}\int\frac{\D^3k'}{(2\pi)^3}\biggl[B+\frac{A}{2}\biggl(\frac{{\bf k}'\cdot{({\bf k}-{\bf k}'})}{k'^2}-\frac{({\bf k}\cdot{\bf k}')({\bf k}\cdot({\bf k}-{\bf k}'))}{k^2k'^2}\biggr)\biggr]\zeta({\bf k}')\zeta({\bf k}-{\bf k}')+\cdots, \label{redef}
\end{align}
\end{widetext}
where
\begin{align}
A&:=\frac{H\mathcal{G}_S}{\Theta\mathcal{G}_T}\frac{\partial\Theta}{\partial H}
-\frac{{H}\mathcal{G}_S}{\mathcal{G}_T^2}\frac{\partial{\cal G}_T}{\partial H}
={\rm const},\\
B&:=\frac{H\mathcal{G}_T\mathcal{G}_S}{\Theta\mathcal{F}_S}={\rm const}.
\end{align}
Here we approximated the time derivative of the curvature perturbation
on superhorizon scales as
\begin{align}
\dot\zeta\simeq-\frac{3(1-n)}{n}H\zeta
\end{align}
and ignored
sub-leading contributions
denoted by the ellipsis ($\cdots$).

The bispectrum $B_\zeta$ is defined by
\begin{align}
\langle\hat\zeta({\bf k}_1)\hat\zeta({\bf k}_2)\hat\zeta({\bf k}_3)
\rangle=(2\pi)^3\delta\left({\bf k}_1+{\bf k}_2+{\bf k}_3\right)B_\zeta,
\end{align}
where we write
\begin{align}
B_\zeta:=(2\pi)^4\frac{\mathcal{P}_\zeta^2}{k_1^3k_2^3k_3^3}\mathcal{A}_{\rm total},
\end{align}
and evaluate the amplitude ${\cal A}_{\rm total}$.
In our setup, $\mathcal{A}_{\rm total}$ reads
\begin{align}
\mathcal{A}_{\rm total}=\mathcal{A}_{\rm original}+\mathcal{A}_{\rm redefine},
\end{align}
where $\mathcal{A}_{\rm original}$ and $\mathcal{A}_{\rm redefine}$ are
the contributions respectively from the interaction Hamiltonian
and from the field redefinition~(\ref{redef}):
\begin{widetext}
\begin{align}
\mathcal{A}_{\rm original}&=\frac{1}{8}\biggl[\biggl(\frac{9(1-n)}{n}\Lambda_1-\Lambda_2+\frac{\Lambda_5}{2}\biggr)\sum_i{k_i^3}+\frac{\Lambda_6}{2}\sum_{i\neq j}k_i^2k_j\notag\\
&\quad +\frac{1}{2k_1^2k_2^2k_3^2}
\biggl(\Lambda_6\sum_i{k_i^9}-(\Lambda_5+\Lambda_6)\sum_{i\neq j}k_i^7k_j^2 -\Lambda_6\sum_{i\neq j}k_i^6k_j^3+(\Lambda_5+\Lambda_6)\sum_{i\neq j}k_i^5k_j^4\biggr)\biggr],\\
\mathcal{A}_{\rm redefine}&=\frac{3}{8}\frac{(1-n)}{n}
\Biggl[(A-4B)\sum_ik_i^3+\frac{A}{4}\sum_{i\neq{j}}k_i^2k_j-\frac{A}{4}\frac{1}{k_1^2k_2^2k_3^2}
\biggl(\sum_{i\neq{j}}k_i^7k_j^2+\sum_{i\neq{j}}k_i^6k_i^3-2\sum_{i\neq{j}}k_i^5k_j^4\biggr)
\Biggr].
\end{align}
\end{widetext}
One can check that the result of the
calculation of the primordial bispectra involving
the procedure of the field redefinition
is identical to that involving boundary terms in the cubic action
with the linear equation of motion $E_S=0$ being imposed. (See Refs.~\cite{Arroja:2011yj,Rigopoulos:2011eq,Burrage:2011hd}.)
The explicit form of the boundary terms is
given in Appendix~\ref{App.C}.

Based on the above result we
also evaluate the nonlinearity parameter defined as
\begin{align}
f_{\rm NL}(k_1,k_2,k_3)=\frac{10}{3}\frac{\mathcal{A}_{\rm total}}{\sum_ik_i^3}
\end{align}
at
the squeezed limit $(k_1\ll{k_2}={k_3})$,
the equilateral limit $(k_1=k_2=k_3)$,
and the folded limit $(k_1=2k,\ k_2=k_3=k)$.
At these limits, the parameter is given respectively by
\begin{align}
f_{\rm NL}^{\rm local}&=\frac{5}{12}\left[\frac{9(1-n)}{n}\Lambda_1-\Lambda_2+3(A-4B)\frac{1-n}{n}\right],\\
f_{\rm NL}^{\rm equil}&=\frac{5}{12}\biggl[\frac{9(1-n)}{n}\Lambda_1-\Lambda_2+\frac{\Lambda_5}{2}+\frac{\Lambda_6}{2}\notag\\
&\quad\quad\quad+\left(\frac{9}{2}A-12B\right)\frac{1-n}{n}\biggr],\\
f_{\rm NL}^{\rm folded}&=\frac{5}{12}\biggl[\frac{9(1-n)}{n}\Lambda_1-\Lambda_2-\frac{8}{5}\Lambda_5+\frac{16}{5}\Lambda_6
\notag \\ &\quad \quad\quad
-12B\frac{1-n}{n}\biggr].
\end{align}
(Here we denoted the nonlinearity parameter at the squeezed limit as
$f_{\rm NL}^{\rm local}$.)

In the case of the matter contracting models within the k-essence theory, these are written as
\begin{align}
f_{\rm NL}^{\rm local}&=\frac{5}{12}\left[-6c_s^2\frac{\lambda}{\mpl^2H^2}-\frac{15}{2}+\frac{9}{4c_s^2}\right],\\
f_{\rm NL}^{\rm equil}&=\frac{5}{12}\left[-6c_s^2\frac{\lambda}{\mpl^2H^2}-\frac{15}{2}+\frac{87}{32c_s^2}\right],\\
f_{\rm NL}^{\rm folded}&=\frac{5}{12}\left[-6c_s^2\frac{\lambda}{\mpl^2H^2}-\frac{15}{2}+\frac{24}{5c_s^2}\right],
\end{align}
where $\lambda:=X^2G_{2XX}+(2/3)X^3G_{2XXX}$.
These results reproduce those in~\cite{Li:2016xjb,Cai:2009fn}.
In order for these nonlinearity parameters to be $\lesssim{\cal O}(1)$,
one requires $c_s^2={\cal O}(1)$.
In the context of k-essence, this leads to $r > {\cal O}(10)$,
which is ruled out.
Instead one may take $c_s^2\ll 1$ to have $r<0.064$,
but then the nonlinearity parameters are
too large to be consistent with observations:
\begin{align}
f_{\rm NL}^{\rm local},\ f_{\rm NL}^{\rm equil},\ f_{\rm NL}^{\rm folded}\sim\frac{1}{c_s^2}=\left(\frac{24}{r}\right)^2>\mathcal{O}(10^5),
\end{align}
indicating that any matter bounce models in the k-essence theory are excluded.
(Observational constraints are given by
$f_{\rm NL}^{\rm local}=0.8\pm5.0$
and
$f_{\rm NL}^{\rm equil}=-4\pm43$~\cite{Akrami:2019izv}.)

Although small $r$ is incompatible with
small scalar non-Gaussianity in the k-essence theory,
this is not always the case in the Horndeski theory.
Thanks to a sufficient number of independent functions,
one can make $r$ small while retaining $A$, $B$, and $\Lambda_i$
less than ${\cal O}(1)$. We will discuss this point in more detail
in the next subsection.

\subsection{Example}

Let us consider a concrete Lagrangian characterized by
\begin{align}
&G_2=\frac{\mpl^2}{\mu^2}e^{-2\phi/\mu}g_2(Y),
\quad
G_3=\frac{\mpl^2}{\mu}g_3(Y),
\notag \\ &
G_4=\frac{\mpl^2}{2},\quad G_5=0,
\end{align}
where $Y:=Xe^{2\phi/\mu}$.
We seek for
a solution of the matter-dominated contracting universe,
$H=2/3t$, with a time-dependent scalar field,
\begin{align}
\phi = \mu \ln (-M t).
\end{align}
It then follows that $Y= \bar Y:= M^2\mu^2/2=\,$const.
This indeed solves the background equations provided that
the functions $g_2(Y)$ and $g_3(Y)$ satisfy
\begin{align}
g_2(\bar Y)&=0,\label{condition1}
\\
g_2'(\bar Y)+2\bar Y g_3'(\bar Y)
&=\frac{4}{3},\label{condition2}
\end{align}
where a prime in this subsection denotes differentiation with respect to $Y$.

Let us further impose that
\begin{align}
\bar Yg_3'(\bar Y)&=\delta_1-1,\label{condition3}
\\
\bar Y\left[g_2''(\bar Y)+2\bar Yg_3''(\bar Y)\right]&=
\frac{1}{3}\left(
21\delta_1+5\delta_2-14\right),\label{condition4}
\end{align}
where $\delta_1$ and $\delta_2$ are some small positive numbers,
$\delta_1\sim \delta_2\ll 1$.
We then have
\begin{align}
{\cal F}_S\simeq \frac{3}{5}\delta_1\mpl^2,
\quad
{\cal G}_S\simeq \frac{3}{5}\delta_2\mpl^2,
\end{align}
and a small tensor-to-scalar ratio can be obtained,
$r=16\delta_1^{3/2}\delta_2^{-1/2}\ll 1$,
while $c_s^2=\delta_1/\delta_2={\cal O}(1)$,
which cannot be achieved in the k-essence theory.

A would-be dangerous contribution to $f_{\rm NL}$ comes from $\Lambda_1$:
\begin{align}
\Lambda_1= -\frac{4}{25\delta_2}
\left[
8+\bar Y^2\left( g_2'''-12g_3''+2\bar Y g_3''' \right)
\right]+{\cal O}(1).
\end{align}
This can be made safe if one requires
\begin{align}
\bar Y^2\left[ g_2'''(\bar Y)-12g_3''(\bar Y)+2\bar Y g_3'''(\bar Y)
 \right]=\delta_3-8,\label{condition5}
\end{align}
where $\delta_3(\lesssim \delta_1)$ is another small number.
All the other terms give at most ${\cal O}(1)$ contributions.

To sum up, by introducing the functions $g_2(Y)$ and $g_3(Y)$
satisfying the conditions~\eqref{condition1},
\eqref{condition2}, \eqref{condition3}, \eqref{condition4}, and~\eqref{condition5},
one has $r\ll 1$ and $f_{\rm NL}\lesssim 1$ simultaneously.
Clearly, this is indeed possible. One can thus circumvent the no-go theorem presented in~\cite{Li:2016xjb}
by appropriately choosing the functions in the Lagrangian
which is more general than the k-essence theory.

\subsection{Tensor Perturbations}

The three-point correlation function including interactions
among different polarization modes of tensor perturbations
can be computed from
\begin{align}
&\langle\hat\xi^{s_1}({\bf k}_1)\hat\xi^{s_2}({\bf k}_2)\hat\xi^{s_3}({\bf k}_3)\rangle\notag\\
&\ =-i\int^{t_b}_{-\infty}dt'\langle[\hat\xi^{s_1}(t_b,{\bf k}_1)
\hat\xi^{s_2}(t_b,{\bf k}_2)\hat\xi^{s_3}(t_b,{\bf k}_3),H_{{\rm int}}(t')]\rangle,
\end{align}
where $\hat\xi^{s}({\bf k}):=\hat h_{ij}({\bf k})e_{ij}^{*(s)}({\bf k})$. The interaction Hamiltonian, $H_{\rm int}$, is given by
\begin{align}
H_{\rm int}=-\int{\D^3x}\mathcal{L}_h^{(3)},
\end{align}
where~\cite{Gao:2011vs}
\begin{align}
\mathcal{L}_h^{(3)}=a^3&\biggl[\frac{\mu}{12}
{\dot h}_{ij}{\dot h}_{jk}{\dot h}_{ki}+\frac{\mathcal{F}_T}{4a^2}\left(h_{ik}h_{jl}
-\frac{1}{2}h_{ij}h_{kl}\right)h_{ij,kl}\biggr],
\end{align}
with $\mu:=-(1/2)\partial\mathcal{G}_T/\partial{H}$
which scales as $\mu\sim(-t)^{3+2\alpha}$,
as seen from Eq.~(\ref{GF}).
The first term, $\dot{h}^3$,
is the new contribution due to $G_{5X}\neq 0$,
while
the second one, which is of the form $h^2\partial^2{h}$,
is identical to the corresponding term in general relativity
except for the overall normalization.
We attach the label ``new'' (respectively, ``GR'') to
the quantities associated with the former (respectively, latter) interaction.

Similarly to the case of the curvature perturbation,
the bispectrum is defined by
\begin{align}
\langle\hat\xi^{s_1}({\bf k}_1)\hat\xi^{s_2}({\bf k}_2)\hat\xi^{s_3}({\bf k}_3)\rangle=&(2\pi)^3\delta\left({\bf k}_1+{\bf k}_2+{\bf k}_3\right)\notag\\
&\quad\times\left({\mathcal{B}}^{s_1s_2s_3}_{(\rm new)}+{\mathcal{B}}^{s_1s_2s_3}_{(\rm GR)}\right),
\end{align}
where
\begin{align}
{\mathcal{B}}^{s_1s_2s_3}_{(\rm new)}&=
\left(2\pi\right)^4\frac{\mathcal{P}_h^2}{k_1^3k_2^3k_3^3}
{\mathcal{A}}^{s_1s_2s_3}_{(\rm new)},\\
{\mathcal{B}}^{s_1s_2s_3}_{(\rm GR)}&=
\left(2\pi\right)^4\frac{\mathcal{P}_h^2}{k_1^3k_2^3k_3^3}
{\mathcal{A}}^{s_1s_2s_3}_{(\rm GR)},
\end{align}
and we evaluate the amplitudes $\mathcal{A}^{s_1s_2s_3}_{(\rm new)}$
and $\mathcal{A}^{s_1s_2s_3}_{(\rm new)}$.
In our setup we obtain
\begin{align}
{\mathcal{A}}^{s_1s_2s_3}_{({\rm new})}=&
\left.\frac{3}{16}\frac{1-n}{n}\frac{H\mu}{\mathcal{G}_{T}}\right|_{t=t_b}
F(s_1k_1,s_2k_2,s_3k_3)\sum_i{k_i^3}, \label{nongcont1}\\
{\mathcal{A}}^{s_1s_2s_3}_{({\rm GR})}=&-\frac{1}{128}
c_t^2\eta_b^2(s_1k_1+s_2k_2+s_3k_3)^2\notag\\
&\quad\times F(s_1k_1,s_2k_2,s_3k_3)\sum_i{k_i^3},\label{nongcont2}
\end{align}
with
\begin{align}
F(x,y,z):=&\frac{1}{64}\frac{1}{x^2y^2z^2}(x+y+z)^3\notag\\
&\ \times(x-y+z)(x+y-z)(x-y-z).
\end{align}
Figures 1 and 2 show
that both ${\mathcal{A}}^{+++}_{(\rm new)}$
and ${\mathcal{A}}^{+++}_{(\rm GR)}$ have peaks at the squeezed limit.
Note that ${\cal A}_{\rm (GR)}^{s_1s_2s_3}$
has a specific scale-dependence $c_t^2k_i^2\eta_b^2$.
This has been obtained in the context of matter bounce cosmology
driven by a scalar field minimally coupled to gravity~\cite{Chowdhury:2015cma}.
However, this factor
makes the detection more challenging~\cite{Kothari:2019yyw}.

\begin{figure}[htb]
\begin{center}
\includegraphics[width=70mm]{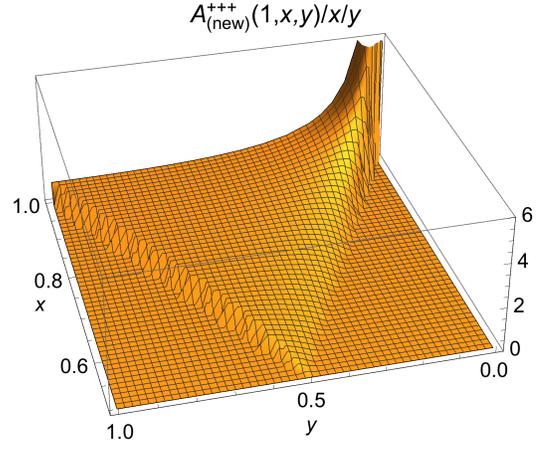}
\caption{$\mathcal{A}^{+++}_{({\rm new})}\left(1,k_2/k_1,k_3/k_1\right)(k_1/k_2)(k_1/k_3)$
as a function of $x=k_2/k_1$ and $y=k_3/k_1$. We take
$n=2/3$ and $H\mu/\mathcal{G}_T|_{t_b}=1$.
The plot is normalized to $1$ for the equilateral configuration, $x=1=y$.}
\end{center}
\end{figure}

\begin{figure}[htb]
\begin{center}
\includegraphics[width=70mm]{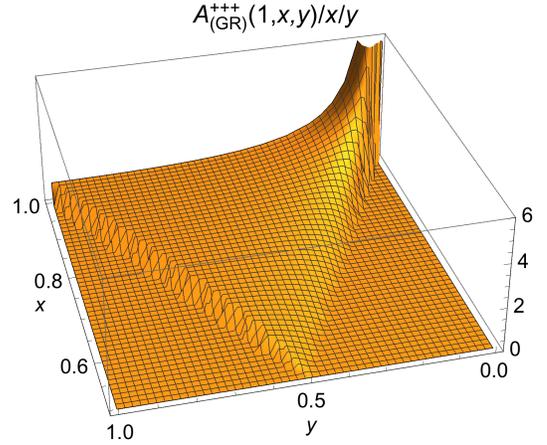}
\caption{$\mathcal{A}^{+++}_{({\rm GR})}\left(1,k_2/k_1,k_3/k_1\right)(k_1/k_2)(k_1/k_3)$
 as a function of $x=k_2/k_1$ and $y=k_3/k_1$. We take
$c_t^2\eta_b^2(k_1+k_2+k_3)^2/128=10^{-6}$. The plot
is normalized to $1$ for the equilateral configuration, $x=1=y$.}
\end{center}
\end{figure}

Now let us compare the above results with the prediction from
generalized G-inflation~\cite{Kobayashi:2011nu}.
The amplitudes of non-Gaussianities
of tensor perturbations in (quasi-de Sitter) inflation
are given by~\cite{Gao:2011vs}
\begin{align}
{\mathcal{A}}^{s_1s_2s_3}_{({\rm new})}=&\frac{H\mu}{4\mathcal{G}_{T}}\frac{k_1^2k_2^2k_3^2}{K^3}F(s_1k_1,s_2k_2,s_3k_3),\\
{\mathcal{A}}^{s_1s_2s_3}_{({\rm GR})}=&\frac{\mathcal{A}}{2}(s_1k_2+s_2k_2+s_3k_3)^2F(s_1k_1,s_2k_2,s_3k_3),
\label{GRampng}
\end{align}
where
\begin{align}
\mathcal{A}=-\frac{K}{16}\left[1-\frac{1}{K^3}\sum_{i\neq{j}}k_i^2k_j-4\frac{k_1k_2k_3}{K^3}\right].
\end{align}
Let us first look at their shapes.
As shown in~\cite{Gao:2011vs},
$\mathcal{A}^{+++}_{\rm (new)}$ of inflation models
has a peak at the equilateral limit.
This is in contrast with the case of contracting models.
On the other hand,
$\mathcal{A}^{+++}_{\rm (GR)}$ has a peak at the squeezed limit
both in inflation and contracting models.
Therefore, the detection of the equilateral-type tensor non-Gaussianities
would
rule out our contracting models.

Next, let us compare the amplitudes.
Squeezed tensor non-Gaussianity from inflation
has the fixed amplitude, as Eq.~\eqref{GRampng}
is independent of the functions in the Horndeski action.
This is not the case for squeezed non-Gaussianity
from contracting models, as is clear from Eqs.~\eqref{nongcont1} and~\eqref{nongcont2},
whichever is dominant.

Finally, notice that
the non-Gaussian amplitudes~\eqref{nongcont1} and~\eqref{nongcont2}
agree with those obtained in a kind of non-attractor inflation models,
where tensor perturbations grow on superhorizon scales during inflation
due to non-attractor dynamics of the non-minimally coupled inflaton~\cite{Ozsoy:2019slf}.
This is because both our contracting models and the
non-attractor phase of inflation are conformally equivalent to
the matter-dominated contracting scenario.

\section{Summary}

In this paper, we
have studied the
primordial power spectra and the bispectra of scalar and tensor perturbations
generated during a general contracting phase
in the Horndeski theory.
It can be shown that under certain conditions
the power spectra of scalar and tensor perturbations are scale invariant.
We have
found that the previous no-go theorem~\cite{Li:2016xjb}
prohibiting
the simultaneous realization of
small tensor-to-scalar ratio and small scalar non-Gaussianity
in matter bounce cosmology driven by a k-essence field
no longer
holds in more general setups.
A concrete example with small $r$ and small $f_{\rm NL}$
has been presented.

Then, we have found that the non-Gaussianities
of tensor perturbations from the contracting universes have two specific features
which are in contrast with the predictions from generalized G-inflation.
First, our contracting models predict only squeezed-type non-Gaussianities,
while inflation can in principle generate
both squeezed- and equilateral-type ones.
Second, the squeezed-type non-Gaussian amplitude from inflation
is model-independently fixed,
while that from the contracting scenario is model-dependent.
We thus conclude that our general bounce model can be distinguished
from generalized G-inflation by combining
the information of
the non-Gaussian amplitudes and shapes.
It would be interesting to investigate the possibility to detect the non-Gaussian signatures predicted from the general bounce model through the B-mode polarization,
as argued in Refs.~\cite{Kothari:2019yyw,Tahara:2017wud}.

\acknowledgments
We would like to thank Jerome Quintin for helpful correspondence. We thank Shuichiro Yokoyama for fruitful discussions. SH thanks Sakine Nishi and Kazufumi Takahashi for instructing him how to use Mathematica for the calculation of perturbations. 
The work of SA was supported by the JSPS Research Fellowships for Young Scientists
No.~18J22305. The work of SH was supported by the JSPS Research Fellowships for Young Scientists No.~17J04865.
The work of TK was supported by
MEXT KAKENHI Grant Nos.~JP15H05888, JP17H06359, JP16K17707, and JP18H04355.


\appendix
\begin{widetext}

\section{Background Equations}\label{App.A}

For a flat FLRW universe the gravitational field equations read~\cite{Kobayashi:2011nu}
\begin{align}
{\cal E}:=\sum_{i=2}^5{\cal E}_i=0,
\quad
{\cal P}:=\sum_{i=2}^5{\cal P}_i=0,
\end{align}
where
\begin{align}
\mathcal{E}_2&=2XG_{2X}-G_2,\\
\mathcal{E}_3&=6X\dot{\phi}HG_{3X}-2XG_{3\phi},\\
\mathcal{E}_4&=-6H^2G_4+24H^2X(G_{4X}+XG_{4XX})-12HX\dot{\phi}G_{4\phi{X}}-6H\dot{\phi}G_{4\phi},\\
\mathcal{E}_5&=2H^3X\dot{\phi}(5G_{5X}+2XG_{5XX})-6H^2X(3G_{5\phi}+2XG_{5\phi{X}}),
\end{align}
and
\begin{align}
\mathcal{P}_2&=G_2,\\
\mathcal{P}_3&=-2X(G_{3\phi}+\ddot{\phi}G_{3X}),\\
\mathcal{P}_4&=2(3H^2+2\dot{H})G_4-12H^2XG_{4X}-4H\dot{X}G_{4X}-8\dot{H}XG_{4X}\notag\\
&\quad-8HX\dot{X}G_{4XX}+2(\ddot{\phi}+2H\dot{\phi})G_{4\phi}+4XG_{4\phi\phi}+4X(\ddot{\phi}-2H\dot{\phi})G_{4\phi{X}},\\
\mathcal{P}_5&=-2X(2H^3\dot{\phi}+2H\dot{H}\dot{\phi}+3H^2\ddot{\phi})G_{5X}
-4H^2X^2\ddot{\phi}G_{5XX}\notag\\
&\quad+4HX(\dot{X}-HX)G_{5\phi{X}}+2\left[2(HX)^{{\boldsymbol \cdot}}+3H^2X\right]G_{5\phi}+4HX\dot{\phi}G_{5\phi\phi}.
\end{align}
The scalar-field equation follows from the above two equations.
\end{widetext}

\section{Effects of Spatial Curvature and Anisotropies on a General Contracting Background}\label{App.B}
In the simple, standard case of a scalar
field minimally coupled to gravity,
spatial curvature and anisotropies in the Friedmann
and evolution equations evolve
in proportion to $a^{-2}$ and $a^{-6}$, respectively.
As a result, it has been known that a contracting universe is
plagued with the instability associated with large anisotropies~\cite{Belinsky:1970ew}.
Some resolutions of the problem have been proposed so far.
See, e.g., Refs.~\cite{Khoury:2001wf,Lehners:2008vx,Cai:2013vm,Qiu:2013eoa,Lin:2017fec}.
However, the impact of spatial curvature and anisotropies
has not been clear yet
in more general cases where the scalar field is nonminimally coupled to gravity.
Hence,
we investigate the evolution of spatial curvature and anisotropies in
a general contracting background in the Horndeski theory.

First, we investigate the impact of
spatial curvature (denoted hereafter as $\mathcal{K}$).
To do so, we consider
open $(\mathcal{K}<0)$ and closed $(\mathcal{K}>0)$ universes in the Horndeski theory.
In the presence of spatial curvature,
the background equations reduce to~\cite{Nishi:2015pta,Akama:2018cqv}
\begin{align}
\mathcal{E}+\mathcal{E}_{\mathcal{K}}=0,\quad
\mathcal{P}+\mathcal{P}_{\mathcal{K}}=0,
\end{align}
where
\begin{align}
\mathcal{E}_{\mathcal{K}}=-3\mathcal{G}_T\frac{\mathcal{K}}{a^2},\quad
\mathcal{P}_{\mathcal{K}}=\mathcal{F}_T\frac{\mathcal{K}}{a^2}.
\end{align}
It can be seen from the scaling argument that
$\mathcal{E}_{\mathcal{K}}/\mathcal{E},\ %
\mathcal{P}_{\mathcal{K}}/\mathcal{P}\propto(-t)^{2(1-n)}$, which
implies that the relative magnitudes of the curvature terms decrease with time
so that
the effect of the spatial curvature on the background equations can be neglected in our setups.

Next, let us consider the effect of anisotropies on the contracting background
by investigating an anisotropic Kasner universe whose metric is written as
\begin{align}
\D s^2&=-\D t^2+a^2\biggl[e^{2(\beta_++\sqrt{3}\beta_-)}\D x^2
\notag \\ &\quad +e^{2(\beta_+-\sqrt{3}\beta_-)}\D y^2
+e^{-4\beta_+}\D z^2\biggr].
\end{align}
The differences between the expansion rates in different directions,
$\beta_\pm$, obey~\cite{Nishi:2015pta,Tahara:2018orv}
\begin{align}
\frac{\D}{\D t}\left\{a^3\left[\mathcal{G}_T\dot\beta_+-2\mu\left(\dot\beta_+^2-\dot\beta_-^2\right)\right]\right\}&=0, \label{beta+}\\
\frac{\D}{\D t}\left\{a^3\left[\mathcal{G}_T\dot\beta_-+4\mu\dot\beta_+\dot\beta_-\right]\right\}&=0. \label{beta-}
\end{align}
Since we have $\mathcal{O}(\mathcal{G}_T)\gtrsim\mathcal{O}(\mu H)$,
the nonlinear terms can be ignored as long as initially
small anisotropies are considered, $\dot\beta_\pm \ll H$.
Then, these equations can be integrated to give
$\dot\beta_\pm \propto (a^3\mathcal{G}_T)^{-1}\propto(-t)^{-(2+2\alpha+3n)}$.
We thus see that
$\dot\beta_\pm/H\propto (-t)^{-(1+2\alpha+3n)}$,
which decreases with time if $1+2\alpha+3n<0$ and
increases if $1+2\alpha+3n>0$.
The case of $\alpha=-2\ (\nu_s=\nu_t=2/3)$ corresponds to the former,
while $\alpha=1-3n\ (\nu_s=\nu_t=-2/3)$ to the latter.
This result implies the contracting background we
are considering requires some mechanism to evade the
unwanted growth of anisotropies. In the present paper, we simply assume
that the contracting universe enjoys a bounce before
the anisotropies spoil its background evolution.

\begin{widetext}
\section{Cubic Action for Scalar Perturbations in the Horndeski Theory}\label{App.C}

Substituting the perturbed metric~\eqref{pertmet}
into the Horndeski action, expanding it to cubic order in perturbations
and using the background equations, we obtain the cubic action for scalar
perturbations~\cite{Gao:2011qe,DeFelice:2011uc,Gao:2012ib}:
\begin{align}
S^{(3)}_S=\int{{\D}t{\D}^3xa^3}&\biggl[\mathcal{G}_T\biggl(-9\zeta{\dot \zeta}^2+\frac{2\dot{\zeta}}{a^2}\left(\zeta\partial^2\chi+\partial_i\zeta\partial_i\chi\right)+\frac{1}{a^4}\left(\partial_i\chi\right)^2\partial^2\zeta+\frac{1}{2a^4}\zeta\left(\left(\partial^2\chi\right)^2-\left(\partial_i\partial_j\chi\right)^2\right)\biggr)\notag\\
&\ -\mathcal{G}_T\frac{\delta{n}}{a^2}\left(\left(\partial_i\zeta\right)^2+2\zeta\partial^2\zeta\right)+\frac{\mathcal{F}_T}{a^2}\zeta\left(\partial_i\zeta\right)^2+3\Sigma\zeta\delta{n}^2+2\Theta\delta{n}\left(9\zeta\dot{\zeta}-\zeta\partial^2\chi-\partial_i\zeta\partial_i\chi\right)\notag\\
&\ +\mu\left(2{\dot \zeta}^3-\frac{2}{a^2}\partial^2\chi{\dot \zeta}^2+\frac{\dot \zeta}{a^4}\left(\left(\partial^2\chi\right)^2-\left(\partial_i\partial_j\chi\right)^2\right)+4\delta{n}\dot\zeta\frac{\partial^2\zeta}{a^2}-\frac{2\delta{n}}{a^4}\left(\partial^2\zeta\partial^2\chi-\partial_i\partial_j\zeta\partial_i\partial_j\chi\right)\right)\notag\\
&\ +\Gamma\left(3\delta{n}{\dot \zeta}^2-\frac{2}{a^2}\delta{n}\dot{\zeta}\partial^2\chi+\frac{1}{2a^4}\delta{n}\left(\left(\partial^2\chi\right)^2-\left(\partial_i\partial_j\chi\right)^2\right)\right)+\Xi\delta{n}^2\left(\dot{\zeta}-\frac{\partial^2\chi}{3a^2}\right)\notag\\
&\ +\left(\Gamma-\mathcal{G}_T\right)\frac{\delta{n}^2}{a^2}\partial^2\zeta-\frac{1}{3}\left(\Sigma+2X\Sigma_X+H\Xi\right)\delta{n}^3\biggr].
\end{align}
From the first-order constraint equations we have
\begin{align}
\delta{n}&=\frac{\mathcal{G}_T}{\Theta}\dot\zeta,\\
\chi&=\frac{1}{a\mathcal{G}_T}\left(a^3\mathcal{G}_S\psi-\frac{a\mathcal{G}_T^2}{\Theta}\zeta\right),
\end{align}
where $\partial^2\psi=\dot\zeta$.
Substituting these solutions into the cubic action, we obtain
\begin{align}
S_\zeta^{(3)}&=\int{{\D}t{\D}^3x}a^3
{\cal G}_S
\biggl\{\frac{\Lambda_1}{H}
\dot\zeta^3+\Lambda_2\zeta\dot\zeta^2+\Lambda_3\zeta\frac{\left(\partial_i\zeta\right)^2}{a^2}
+\frac{\Lambda_4}{H^2}
\dot\zeta^2\frac{\partial^2\zeta}{a^2}
+\Lambda_5\dot\zeta\partial_i\zeta\partial_i
\psi+\Lambda_6\partial^2\zeta\left(\partial_i\psi\right)^2\notag\\
&\quad\quad\quad\quad\quad\quad\quad
+\frac{\Lambda_7}{H^2}
\frac{1}{a^4}\left[\partial^2\zeta\left(\partial_i\zeta\right)^2-\zeta\partial_i\partial_j\left(\partial_i\zeta\partial_j\zeta\right)\right]
+\frac{\Lambda_8}{H}\frac{1}{a^2}\left[\partial^2\zeta\partial_i\zeta\partial_i\psi-\zeta\partial_i\partial_j\left(\partial_i\zeta\partial_j\psi\right)\right]\biggr\}\notag\\
&\quad+\int{{\D}t{\D}^3x}F(\zeta)E_S, \label{cubic}
\end{align}
where
\begin{align}
\Lambda_1&=H
\biggl[\frac{\mathcal{G}_T}{\Theta}\left(\frac{\mathcal{G}_S}{\mathcal{F}_S}+3\frac{\mathcal{G}_T}{\mathcal{G}_S}-1\right)+\frac{\Xi \mathcal{G}_T}{3\Theta^2}\left(3\frac{\mathcal{G}_T}{\mathcal{G}_S}
-1\right)+2\mu\left(\frac{1}{\mathcal{G}_S}
-\frac{1}{\mathcal{G}_T}\right)+\frac{\Gamma}{\Theta}\left(3\frac{\mathcal{G}_T}{\mathcal{G}_S}-2\right) \notag \\
&\quad\quad+\frac{2}{3}\frac{\mathcal{G}_T^3}{\Theta^3\mathcal{G}_S}\left(\Sigma-X\Sigma_X\right)-\frac{H}{3}\frac{\mathcal{G}_T^3\Xi}{\Theta^3\mathcal{G}_S}\biggr],\\
\Lambda_2&=3-\frac{H\mathcal{G}_T\mathcal{G}_S}{\mathcal{F}_S\Theta}\left(3-g_T+f_S+f_\Theta\right),\\
\Lambda_3&=
\frac{\mathcal{F}_T}{\mathcal{G}_S}+\frac{H\mathcal{G}_T}{\Theta}\left(1+g_T+g_S-f_\Theta\right)-\frac{H\mathcal{G}_T^2}{\mathcal{G}_S\Theta}\left(1+2g_T-f_\Theta\right),\\
\Lambda_4&=H^2
\left[\frac{\Xi}{3}\frac{\mathcal{G}_T^3}{\mathcal{G}_S\Theta^3}+6\mu\frac{\mathcal{G}_T}{\mathcal{G}_S\Theta}+\left(3\Gamma-\mathcal{G}_T\right)\frac{\mathcal{G}_T^2}{\mathcal{G}_S\Theta^2}\right],\\
\Lambda_5&=-\frac{1}{2}\frac{\mathcal{G}_S}{\mathcal{G}_T}-\frac{H}{2}\frac{\Gamma\mathcal{G}_S}{\mathcal{G}_T\Theta}\left(3+g_T-f_\Gamma+f_\Theta\right)-
\mu{H}\frac{\mathcal{G}_S}{\mathcal{G}_T^2}\left(3+2g_T-f_\mu\right)
,\\
\Lambda_6&=
\frac{3}{4}\frac{\mathcal{G}_S}{\mathcal{G}_T}-\frac{\mathcal{G}_S}{4\mathcal{G}_T}\frac{\Gamma{H}}{\Theta}\left(3+g_T-f_\Gamma+f_\Theta
\right)-\mu{H}\frac{\mathcal{G}_S}{\mathcal{G}_T^2}\left(\frac{3}{2}+g_T
-\frac{1}{2}f_\mu\right),\\
\Lambda_7&=\frac{H^2}{6}\biggl[\frac{\mathcal{G}_T^3}{\mathcal{G}_S\Theta^2}-\frac{H\Gamma\mathcal{G}_T^3}{\mathcal{G}_S\Theta^3}\left(1-3g_T+3f_\Theta-f_\Gamma+3\frac{\Theta\mathcal{F}_S}{H\mathcal{G}_T^2}\right)
-6\mu{H}\frac{\mathcal{G}_T^2}{\mathcal{G}_S\Theta^2}\left(1-2g_T-f_\mu+2f_\Theta+2\frac{\Theta\mathcal{F}_S}{H\mathcal{G}_T^2}\right)\biggr],\\
\Lambda_8&=H
\biggl[-\frac{\mathcal{G}_T}{\Theta}+\frac{\mu{H}}{\Theta}\left(4+2f_\Theta-2f_\mu+2\frac{\Theta\mathcal{F}_S}{H\mathcal{G}_T^2}\right)+H\frac{\Gamma\mathcal{G}_T}{\Theta^2}\left(1-\frac{1}{2}g_T
-\frac{1}{2}f_\Gamma+f_\Theta+\frac{\Theta\mathcal{F}_S}{H\mathcal{G}_T^2}\right)\biggr],\\
F(\zeta)&=-\frac{\mathcal{G}_T\mathcal{G}_S}{\Theta\mathcal{F}_S}\zeta\dot\zeta-
\frac{1}{2}\left(\frac{\Gamma\mathcal{G}_S}{\Theta\mathcal{G}_T}+2\mu\frac{\mathcal{G}_S}{\mathcal{G}_T^2}\right)
\left(\partial_i\zeta\partial_i\psi-\partial^{-2}\partial_i\partial_j\left(\partial_i\zeta\partial_j\psi\right)\right)
\notag\\&\quad\quad
+\frac{1}{4a^2}\left(\frac{\Gamma\mathcal{G}_T}{\Theta^2}+\frac{4\mu}{\Theta}\right)\left(\left(\partial_i\zeta\right)^2-\partial^{-2}\partial_i\partial_j\left(\partial_i\zeta\partial_j\zeta\right)\right),\\
E_S&=-2\left[\partial_t\left(a^3\mathcal{G}_S\dot\zeta\right)-a\mathcal{F}_S\partial^2\zeta\right].
\end{align}
Here we defined
\begin{align}
\Theta&:=-\dot{\phi}XG_{3X}+2HG_4-8HXG_{4X}\notag-8HX^2G_{4XX}+\dot{\phi}G_{4\phi}+2X\dot{\phi}G_{4\phi{X}}\notag\\&
\quad-H^2\dot{\phi}(5XG_{5X}+2X^2G_{5XX})+2HX(3G_{5\phi}+2XG_{5\phi{X}}), \label{Theta-ap}\\
\Sigma&:=XG_{2X}+2X^2G_{2XX}+12H\dot{\phi}XG_{3X}+6H\dot{\phi}X^2G_{3XX}-2XG_{3\phi}-2X^2G_{3\phi{X}}-6H^2G_4\notag\\ &
\quad +6\bigl[H^2(7XG_{4X}+16X^2G_{4XX}+4X^3G_{4XXX})-H\dot{\phi}(G_{4\phi}+5XG_{4\phi{X}}+2X^2G_{4\phi{X}X})\bigr]\notag\\ &
\quad +2H^3\dot{\phi}\left(15XG_{5X}+13X^2G_{5XX}+2X^3G_{5XXX}\right)-6H^2X(6G_{5\phi}+9XG_{5\phi{X}}+2X^2G_{5\phi{X}X}), \label{Sigma-ap}\\
\Gamma&:=2G_4-8XG_{4X}-8X^2G_{4XX}-2H\dot\phi(5XG_{5X}+2X^2G_{5XX})+2X(3G_{5\phi}+2XG_{5\phi X}), \label{Gamma-ap}\\
\Xi&:=12\dot\phi{X}G_{3X}+6\dot\phi{X^2}G_{3XX}-12HG_4+6\biggl[2H(7XG_{4X}+16X^2G_{4XX}+4X^3G_{4XXX})-\dot\phi(G_{4\phi}\notag\\
&\quad+5XG_{4\phi{X}}+2X^2G_{4\phi{XX}})\biggr]+90H^2\dot\phi{X}G_{5X}+78H^2\dot\phi{X^2}G_{5XX}+12H^2\dot\phi{X^3}G_{5XXX}\notag\\
&\quad-12HX(6G_{5\phi}+9XG_{5\phi X}+2X^2G_{5\phi XX}), \label{Xi-ap}
\end{align}
and
\begin{align}
g_T&=\frac{\dot{\mathcal{G}}_T}{H\mathcal{G}_T},\quad
g_S=\frac{\dot{\mathcal{G}}_S}{H\mathcal{G}_S},\quad
 f_S=\frac{\dot{\mathcal{F}}_S}{H\mathcal{F}_S},\quad
 f_\Theta=\frac{\dot\Theta}{H\Theta},\quad
  f_\Gamma=\frac{\dot\Gamma}{H\Gamma},\quad  f_\mu=\frac{\dot\mu}{H\mu}.
\end{align}
Note that we can write the Eqs.~(\ref{Gamma-ap}), (\ref{Xi-ap}) as
\begin{align}
\Gamma=\frac{\partial \Theta}{\partial H},\quad
\Xi=\frac{\partial\Sigma}{\partial H}.
\end{align}
It is therefore natural to assume that
these quantities scale as
\begin{align}
\Gamma \sim (-t)^{2+2\alpha},
\quad
\Xi\sim (-t)^{1+2\alpha}.
\end{align}

In Eq.~(\ref{cubic}), we neglected some boundary terms
having the form of a total time derivative.
They are given by
\begin{align}
S_B&=\int\D t\D^3x\frac{\D}{\D t}\biggl[-a^3\frac{\mathcal{G}_T\mathcal{G}_S^2}{\Theta\mathcal{F}_S}\zeta\dot\zeta^2
+a^3\frac{\mathcal{G}_S^2}{2\mathcal{G}_T^2}\left(2\mu+\frac{\Gamma\mathcal{G}_T}{\Theta}\right)\left(\zeta\dot\zeta^2-\zeta(\partial_i\partial_j\psi)^2\right)
\notag \\ & \quad \quad\quad\quad\quad\quad
-\frac{a\mathcal{G}_S}{2\Theta}\left(4\mu+\frac{\Gamma\mathcal{G}_T}{\Theta}\right)\left(\zeta\dot\zeta\partial^2\zeta-\zeta\partial_i\partial_j\psi\partial_i\partial_j\zeta\right)
+\frac{9a^3}{2}(A_3-2H\mathcal{G}_T-2\mu H^2)\zeta^3
\notag \\ & \quad \quad\quad\quad\quad\quad
+a\left(\frac{\mathcal{G}_T^2}{\Theta}-B_5\right)\zeta(\partial_i\zeta)^2-\frac{\mathcal{G}_T^2}{6a\Theta^2}\left(6\mu+\frac{\Gamma\mathcal{G}_T}{\Theta}\right)\left(\zeta(\partial_i\partial_j\zeta)^2-\zeta(\partial^2\zeta)^2\right)\biggr],
\end{align}
where
\begin{align}
A_3&=-\int^XG_{3X'}\sqrt{2X'}\D X'-2\sqrt{2X}G_{4\phi},\\
B_5&=-\int^XG_{5X'}\sqrt{2X'}\D X'.
\end{align}
\end{widetext}



\end{document}